\documentclass{article}
\usepackage{spconf,amsmath,amssymb,graphicx}
\usepackage{booktabs,cite,microtype,url,etoolbox}
\usepackage{xurl}
\usepackage{hyperref}

\AtBeginEnvironment{thebibliography}{%
    \small
    \setlength{\itemsep}{0pt}
    \setlength{\parskip}{0pt}
}

\title{Multi-sample Synthetic  Supervision for Accent Conversion}

\name{Yangyang Qu \qquad
Michele Panariello \qquad
Massimiliano Todisco \qquad
Nicholas Evans}
\address{EURECOM, Sophia Antipolis, France}
\hypersetup{pdfauthor={Yangyang Qu, Massimiliano Todisco, Nicholas Evans},pdftitle={Multi-Constraint Synthetic Supervision for Direct Accent Conversion}}
\begin{document}
\ninept

\maketitle


\begin{abstract}
Accent conversion (AC) requires changing accent while preserving speaker identity and linguistic content, yet parallel recordings are scarce.
Speech synthesis provides an alternative source of supervision, but generated targets vary in accent realization and source preservation.
We propose a multi-sample synthetic supervision framework that constructs conversion targets by jointly assessing these properties across candidate waveforms for each source--accent condition.
Selected candidates provide associated discrete speech codes, representing linguistic content, prosody, and speaking style, as targets for accent-conditioned autoregressive adaptation.
Compared with using one generated target per training example, our method improves target-accent classification accuracy by 2.49 percentage points, while speaker similarity remains nearly unchanged, and word error rate increases by 0.29 percentage points.
Across six target accents, our method achieves 7.81\% WER and the highest mean listening ratings among evaluated systems.
\end{abstract}

\begin{keywords}
accent conversion, multi-sample supervision, synthetic supervision, stochastic generation, target-accent conditioning
\end{keywords}



\begin{figure*}[t]
\centering
\includegraphics[width=\textwidth]{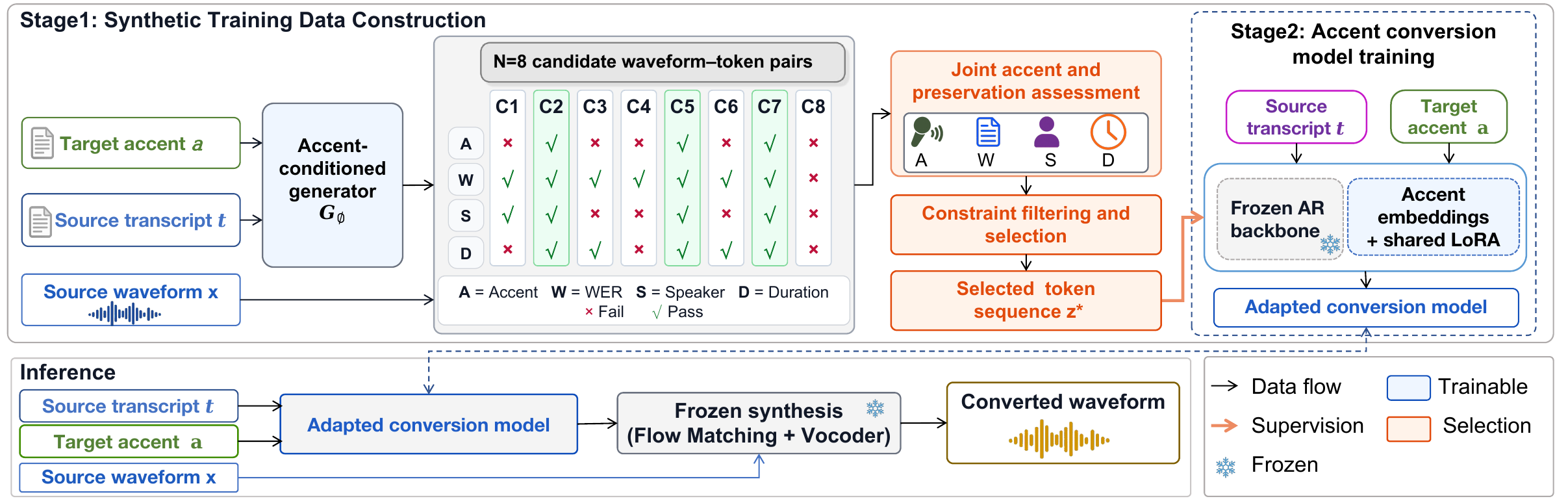}
\caption{
Overview of multi-sample synthetic supervision and accent conversion.
A reference-conditioned generation stage first provides token targets for
adapting the categorical candidate generator $G_\phi$.
The fixed $G_\phi$ then samples $N$ token sequences from the source transcript
$t$ and target-accent label $a$, while frozen acoustic synthesis uses the source
waveform $x$ to produce the paired candidate waveforms.
Waveform-level accent and preservation assessment selects a candidate whose
paired token sequence $z^{*}$ supervises the final converter $p_\theta$.
At inference, $p_\theta$ generates one token sequence from $(t,a)$ and frozen
synthesis combines it with $x$, without target-accent reference speech or
candidate selection.
}
\label{fig:method}
\end{figure*}

\section{Introduction}
\label{sec:introduction}

Accent conversion (AC) aims to modify accent while preserving speaker identity and linguistic content.
Learning this transformation from paired examples ideally requires recordings of the same speaker uttering the same content in different accents.
Such recordings are difficult to collect, motivating the use of synthetic supervision and making the construction of suitable conversion targets a central challenge.

Existing approaches alleviate this data shortage using speech synthesis and transferred representations.
TTS-guided training~\cite{zhou2023ttsguided} and transferred linguistic representations~\cite{chen2024transfer} support AC with non-parallel data.
Other approaches construct synthetic native-like targets~\cite{nguyen2025syntheticgt}, learn conversion through discrete speech tokens~\cite{bai25_interspeech}, or pair synthesized accented speech with natural target recordings~\cite{bai2026cosyaccent}.
Together, these methods show that useful AC supervision can be constructed without directly matched recordings.

Pretrained controllable speech generators such as Vevo~\cite{zhang2025vevo} and Vevo2~\cite{zhang2026vevo2} can also be used to construct synthetic AC supervision.
Because generation is stochastic, repeated sampling under the same source--accent condition can produce realizations with different target-accent and source-preservation characteristics.
This raises a specific supervision question: whether a single sampled target is sufficient, or whether multiple stochastic realizations can provide more effective supervision.

We formulate this question as a multi-sample target-construction problem.
Selection from multiple generations has precedents in best-of-$N$ distillation~\cite{sessa2025bond,zhang2026brts} and ASR-based self-verification for speech synthesis~\cite{asaria2026reliabletts}.
In AC, however, candidate quality involves multiple requirements: the output should exhibit the requested accent change while preserving linguistic content and source-speaker characteristics.
Consequently, transcription correctness alone is insufficient for constructing suitable AC supervision.
We therefore jointly assess accent realization and source preservation at the waveform level, transfer the selection to the associated token sequence, and test the resulting supervision under a matched source--accent population.

We propose a multi-sample synthetic supervision framework in which a fixed categorical accent-conditioned generator produces multiple waveform--token candidates for each source--accent condition.
Here, the tokens are discrete content-style codes from a learned speech representation that encode linguistic content, prosody, and speaking style.
Candidate waveforms are jointly assessed for target-accent realization and source preservation, and the content-style token sequence paired with a selected feasible waveform is used as pseudo-supervision for categorical autoregressive adaptation.
To test whether access to multiple stochastic realizations improves supervision rather than merely expanding the pool of usable training conditions, we compare single- and eight-sample supervision on the same source--accent conditions with identical data splits, initialization within each seed, and optimization protocol.
This controlled comparison isolates the downstream effect of replacing the first generated target with a target selected from multiple stochastic realizations.
Under this matched comparison, eight-sample supervision improves target-accent accuracy by 2.49 percentage points, with nearly unchanged speaker similarity.

Our main contributions are:
\begin{itemize}

\item We formulate synthetic AC supervision as a multi-sample target-construction problem, in which multiple stochastic realizations of the same source--accent condition are jointly assessed for target-accent realization and source preservation.

\item We develop an AC-specific supervision framework that first adapts a pretrained autoregressive backbone to categorical accent control and then transfers waveform-level accent--preservation assessment to paired discrete content-style targets for final autoregressive adaptation.

\item We validate the proposed formulation through a controlled single- versus multi-sample comparison that holds the source--accent population and data split, showing stronger target-accent evidence with only a small WER increase and nearly unchanged speaker similarity.

\end{itemize}
\section{Method}
\label{sec:method}

Given a source waveform $x$, its transcript $t$, and a target-accent label $a$, we aim to generate speech $\hat{x}$ in the requested accent while preserving linguistic content and source-speaker identity:
$
(t,x,a)\rightarrow\hat{x}.
$
The transcript is available during both training and inference.
The autoregressive model is conditioned on $t$ and $a$, whereas $x$ is used only for source timbre conditioning during the subsequent frozen acoustic synthesis stage.
\subsection{Synthetic Training Data Construction}
\label{sec:synthetic_supervision}

\subsubsection{Stochastic Candidate Generation}
\label{sec:candidate_generation}

We construct the accent-conditioned candidate generator $G_\phi$ in two steps.
First, we use reference-conditioned generation to construct the token-level training targets for $G_\phi$.
Let $r_a$ denote a reference utterance representing target accent $a$ and $t_a^r$ its transcript.
Given the source transcript $t$, the target-reference pair $(r_a,t_a^r)$, and source waveform $x$, the pretrained Vevo2 model~\cite{zhang2026vevo2} generates a content-style token sequence for the source linguistic content while the reference provides target-accent and style information.
Frozen acoustic synthesis uses $x$ for source-timbre conditioning.
Selected token sequences from this stage are retained as training targets for the subsequent categorical adaptation.

Second, we use these token targets to train a categorical accent-conditioned autoregressive generator $G_\phi$.
This adaptation replaces target-reference conditioning with categorical accent control, allowing $G_\phi$ to generate content-style tokens directly from a source transcript and target-accent label.
Each target accent is represented by a trainable sequence of 32 accent embeddings, while shared rank-32 LoRA modules~\cite{hu2022lora} adapt the query, key, value, and output projections.
We first optimize the accent embeddings alone and then jointly optimize the accent embeddings and LoRA parameters using autoregressive token cross-entropy.
All remaining autoregressive backbone parameters are frozen.
After development-set checkpoint selection, $G_\phi$ is fixed for multi-sample supervision construction.

Given a source transcript $t$ and target-accent label $a$, the fixed generator $G_\phi$ samples a content-style token sequence $z_i$ conditioned on $(t,a)$.
The content-style tokens are discrete codes from the pretrained content-style representation~\cite{zhang2026vevo2}, representing linguistic content together with prosodic and speaking-style information.
Given $z_i$ and the source waveform $x$, frozen Flow Matching~\cite{zhang2026vevo2} and Vocos~\cite{siuzdak2024vocos} synthesize the corresponding candidate waveform $\tilde{x}_i$, with $x$ providing source-timbre conditioning.
Thus, target-accent reference speech is required only for constructing the initial pseudo-target supervision, whereas the trained generator $G_\phi$ produces candidates using only the source transcript and categorical target-accent label.

Repeating the stochastic generation process under fixed model parameters and decoding settings produces multiple waveform--token realizations for the same source--accent condition.
With $N$ repetitions, we obtain
\[
\mathcal{C}_{N}(t,x,a)
=
\left\{
(\tilde{x}_i,z_i)
\right\}_{i=1}^{N}.
\]
Each pair $(\tilde{x}_i,z_i)$ preserves the correspondence between the synthesized waveform used for quality assessment and the discrete token sequence available for supervision.
The candidate waveform $\tilde{x}_i$ is subsequently evaluated for target-accent realization and source preservation, while its paired token sequence $z_i$ serves as the candidate pseudo-target for final-model training.

\subsubsection{Candidate Assessment and Selection}
\label{sec:target_selection}

We assess each candidate waveform jointly for target-accent evidence, content preservation, speaker preservation, and duration consistency.
Let $q$ denote the fixed development-time accent scorer used for candidate construction.
We define
$
p_i=q(a\mid\tilde{x}_i)
$
and
$
\Delta p_i=p_i-q(a\mid x),
$
and set $h_i=1$ when $q$ predicts the requested accent $a$ for $\tilde{x}_i$.

Let $w_i$ denote candidate WER, $s_i$ the source--candidate ECAPA-TDNN~\cite{desplanques2020ecapa} cosine similarity, and
\[
\rho_i=
\frac{\operatorname{dur}(\tilde{x}_i)}
{\operatorname{dur}(x)}.
\]
A candidate is feasible when
\begin{equation}
w_i\!\leq\!\tau_{\mathrm{WER}},\;
s_i\!\geq\!\tau_{\mathrm{SIM}},\;
\rho_{\min}\!\leq\!\rho_i\!\leq\!\rho_{\max},\;
(h_i\!=\!1\vee\Delta p_i\!\geq\!\tau_{\Delta}).
\label{eq:selection_constraints}
\end{equation}
Conditions with no feasible candidate are excluded from final-model training.

Among feasible candidates, we select the highest-ranked waveform--token pair using lexicographic ordering by decreasing $p_i$, $\Delta p_i$, $h_i$, and $s_i$, followed by increasing $w_i$.
The selected token sequence $z^{*}$ is retained as the pseudo-target for final-model adaptation.

\subsection{Accent Conversion Model Training}
\label{sec:conversion}

After candidate assessment and selection, each retained source--accent condition provides an autoregressive training example
$
(t_b,a_b,z_b^{*}),
$
where $t_b$ is the source transcript, $a_b$ is the categorical target-accent label, and $z_b^{*}$ is the token sequence paired with the selected waveform.
Thus, waveform-level assessment determines which generated content-style token sequence is used as supervision for the final conversion model.

The final converter $p_\theta$ uses the same categorical autoregressive architecture as the candidate generator $G_\phi$, but its parameters are further adapted using the selected supervision.
For each training seed, the converter is initialized from a seed-specific categorical warm-start derived from $G_\phi$, rather than directly from the pretrained backbone.
We then continue adapting the model on the selected token targets $z^{*}$ from the final supervision pool.

As in $G_\phi$, each target accent is represented by a trainable sequence of 32 accent embeddings, and shared rank-32 LoRA modules~\cite{hu2022lora} adapt the query, key, value, and output projections of the autoregressive backbone.
Only the accent embeddings and shared LoRA parameters are optimized, while the remaining autoregressive backbone parameters are frozen.
The autoregressive model therefore learns
$
p_\theta(z\mid t,a),
$
without using the source waveform $x$ as an input to the token predictor.

For a minibatch of $B$ examples, let $M_b$ denote the non-padding positions of the selected token sequence $z_b^{*}$.
We minimize the causal token negative log-likelihood
\begin{equation}
\mathcal{L}_{\mathrm{tok}}
=
-\frac{1}{B}
\sum_{b=1}^{B}
\frac{1}{|M_b|}
\sum_{u\in M_b}
\log p_{\theta}
\left(
z^{*}_{b,u}
\mid
z^{*}_{b,<u},t_b,a_b
\right),
\label{eq:token_nll}
\end{equation}
where $\theta$ denotes the trainable accent-embedding and LoRA parameters.
No additional accent, content, speaker, or preference loss is used during this final adaptation.

\subsection{Inference}
\label{sec:inference}

At inference, given a source transcript $t$ and target-accent label $a$, the trained autoregressive converter generates a single content-style token sequence
$
\hat{z}
\sim
p_{\theta}(z\mid t,a).
$
The source waveform $x$ does not enter the autoregressive converter and is instead used as the source-timbre reference during the subsequent frozen acoustic synthesis stage.
Frozen Flow Matching~\cite{zhang2026vevo2} and Vocos~\cite{siuzdak2024vocos} synthesize the converted waveform $\hat{x}$ from $\hat{z}$ and $x$.
Thus, inference requires only $(t,x,a)$ and produces a single converted waveform without target-accent reference speech, multi-sample candidate generation, or candidate selection.

\begin{table*}[t]
\caption{
Automatic results on AC-Eval3K and listening-test results on the perceptual evaluation subset.
MOS values are descriptive means without uncertainty estimates.
All systems use the same evaluation pipeline, but the external systems are not training- or input-matched to ours.
Higher is better for target-accent classification accuracy, $\Delta p_{\mathrm{tgt}}$, SIM, and MOS; lower is better for WER.
}
\label{tab:main}
\centering
\small
\setlength{\tabcolsep}{5.0pt}
\begin{tabular}{lccccccc}
\toprule
System
& Acc. (\%) $\uparrow$
& $\Delta p_{\mathrm{tgt}}$ (pp) $\uparrow$
& WER (\%) $\downarrow$
& SIM $\uparrow$
& N-MOS $\uparrow$
& S-MOS $\uparrow$
& A-MOS $\uparrow$ \\
\midrule
Vevo-Style
& 32.33 & +13.83 & 14.24 & 0.6737 & 3.440 & 3.429 & 2.940 \\
Vevo1.5
& 18.63 & +1.67 & 31.05 & 0.7657 & 2.857 & 3.179 & 2.417 \\
Vevo2
& 19.13 & +2.16 & 17.01 & \textbf{0.8227} & 3.619 & 3.488 & 2.940 \\
SeedVC-v2
& 42.10 & +20.08 & 17.60 & 0.0910 & 2.464 & 1.786 & 2.440 \\
StarGANv2-VC
& \textbf{45.50} & \textbf{+23.79} & 20.68 & 0.1888 & 1.881 & 1.583 & 1.929 \\
\midrule
\textbf{Ours}
& 30.31 & +11.01 & \textbf{7.81} & 0.7135
& \textbf{3.690} & \textbf{3.560} & \textbf{3.131} \\
\bottomrule
\end{tabular}
\end{table*}
\section{Experimental Setup}
\label{sec:experiments}

\subsection{Data and Implementation}
\label{sec:data_protocol}
\label{sec:implementation}

We evaluate six English target accents: Arabic, Chinese, Hindi, Korean, Spanish, and Vietnamese.
Synthetic-supervision construction uses 29,754 source--accent conditions from 17,649 LibriTTS~\cite{zen2019libritts} utterances from 235 speakers and 12,105 L2-ARCTIC~\cite{zhao2018l2arctic} utterances from 20 speakers, with 4,959 conditions per accent.
The categorical generator $G_\phi$ is first adapted using 29,840 pseudo-target token sequences obtained from the reference-conditioned bootstrap stage in Sec.~\ref{sec:candidate_generation}.

For final supervision construction, the fixed $G_\phi$ generates $N=8$ candidates per condition using top-$k=25$, top-$p=0.8$, temperature $1.0$, and 32 Flow Matching steps.
Candidate retention uses
$\tau_{\mathrm{WER}}=0.08$,
$\tau_{\mathrm{SIM}}=0.60$,
$\rho\in[0.75,1.45]$,
and $\tau_{\Delta}=0.02$.
At least one candidate is feasible for 16,153 conditions, which are partitioned into 15,346 optimization and 807 internal-validation conditions.
A separate 300-condition development set is used for checkpoint selection.

Candidate selection uses a fixed six-way accent probe $q$ based on valid-frame mean-pooled Whisper-small~\cite{radford2023whisper} encoder representations, feature standardization, and balanced logistic regression.
WER and speaker similarity are measured using Whisper-small and ECAPA-TDNN~\cite{desplanques2020ecapa}, respectively.
The final converter is trained for three epochs using seeds 1337, 2027, and 3407.
The learning rates are $5\times10^{-6}$ for LoRA and $10^{-4}$ for the accent embeddings.

We construct AC-Eval3K from 3,000 unique LibriTTS test-clean utterances by assigning each source utterance one target-accent label, yielding 500 conditions for each of the six accents.
The utterances come from 39 speakers, with no source-utterance or speaker overlap with the supervision-construction pool.

\subsection{Compared Systems and Evaluation}
\label{sec:systems_evaluation}
\label{sec:automatic_metrics}

Direct comparison with existing AC-specific systems is limited by differences in task definition and target-accent coverage.
Many closely related methods focus on accent normalization or specific source--target mappings~\cite{chen2024transfer,nguyen2025syntheticgt,bai2026cosyaccent}, and do not provide a released model or protocol covering all six target accents considered here.
Evaluating them on AC-Eval3K would therefore require retraining or modifying their original task configuration.
We instead compare with systems that can generate outputs for all six targets: Vevo-Style~\cite{zhang2025vevo}, Vevo1.5~\cite{li2025overview}, Vevo2~\cite{zhang2026vevo2}, SeedVC-v2~\cite{liu2024seedvc}, and StarGANv2-VC~\cite{li2021starganv2vc}.
All systems are evaluated using the same output-level scoring pipeline, although their training and inference conditions are not matched.

For automatic evaluation, target-accent performance is measured using the PAID-Whisper evaluation framework introduced in~\cite{qu2026phonemeaware}.
We report target-accent classification accuracy (Acc.), defined as the percentage of converted utterances classified as the requested accent, and
$
\Delta p_{\mathrm{tgt}}
=
p(a\mid\hat{x})-p(a\mid x),
$
which measures the source-relative increase in probability assigned to the requested accent.
Content preservation is measured by word error rate (WER) using Whisper-small~\cite{radford2023whisper}, and source-speaker preservation by cosine similarity (SIM) between SpeechBrain ECAPA-TDNN~\cite{desplanques2020ecapa} embeddings of the source and converted speech.
Higher Acc., $\Delta p_{\mathrm{tgt}}$, and SIM indicate stronger target-accent evidence or speaker preservation, whereas lower WER indicates better content preservation.

To isolate the effect of multi-sample supervision, we compare $N=1$ and $N=8$ on the same 6,600 source--accent conditions.
The $N=1$ system uses the first feasible candidate, whereas $N=8$ selects the highest-ranked feasible candidate from eight generations.
Within each seed, both systems use the same data split, initialization, and training protocol.
Target-accent metrics in this matched comparison are measured using the development accent probe $q$.

We further conduct a blinded listening test with 10 listeners on 12 source utterances, comprising one male and one female utterance per target accent.
Listeners rate naturalness (N-MOS), source-speaker similarity (S-MOS), and target-accent similarity (A-MOS) on 1--5 scales.
For A-MOS, listeners are provided with target-accent reference speech only for perceptual evaluation.

Code and audio samples are available at
\url{https://github.com/eurecom-asp/SSAC}
and
\url{https://github.com/eurecom-asp/SSAC_listening_samples}.

\begin{table}[t]
\caption{
Single- versus multi-sample supervision using the same source utterance--target accent training combinations.
Values are means over three training runs.
}
\label{tab:candidate_budget}
\centering
\small
\setlength{\tabcolsep}{4.0pt}
\begin{tabular}{lcccc}
\toprule
Samples
& Acc. (\%) $\uparrow$
& $\Delta p_{\mathrm{tgt}}$ (pp) $\uparrow$
& WER (\%) $\downarrow$
& SIM $\uparrow$ \\
\midrule
$N=1$
& 28.08
& +9.77
& \textbf{7.82}
& \textbf{0.7139} \\
$N=8$
& \textbf{30.57}
& \textbf{+11.70}
& 8.11
& 0.7134 \\
\bottomrule
\end{tabular}
\end{table}


\section{Results}
\label{sec:results}

\subsection{Overall Conversion Performance}
\label{sec:main_results}

Table~\ref{tab:main} summarizes performance across the evaluated systems.
Our system achieves the lowest WER of 7.81\%, with a target-accent classification accuracy of 30.31\% and $\Delta p_{\mathrm{tgt}}=+11.01$ pp.
Vevo-Style, SeedVC-v2, and StarGANv2-VC obtain higher automatic target-accent scores, while Vevo2 achieves the highest SIM.
These results show different accent--content--speaker operating points under the common output-level evaluation rather than dominance by a single system across all metrics.

In the listening test, our system obtains the highest observed mean ratings for naturalness, source-speaker similarity, and target-accent similarity, with N-MOS, S-MOS, and A-MOS of 3.690, 3.560, and 3.131, respectively.

\subsection{Effect of Multi-Sample Supervision}
\label{sec:n_budget}

In the full construction pool, the first generated candidate satisfies the retention requirements for 6,600 of 29,754 source--accent conditions (22.18\%).
With eight stochastic generations, 16,153 conditions (54.29\%) contain at least one feasible candidate.
This increased coverage shows that multiple generations provide more opportunities to construct usable supervision, but does not by itself establish a downstream benefit.

We therefore compare $N=1$ and $N=8$ using the same 6,600 source--accent conditions for which the first candidate is feasible.
For $N=1$, the first candidate provides the supervision target, whereas for $N=8$, the highest-ranked feasible candidate is selected from eight generations.
Within each seed, the two systems use the same data split, categorical warm-start, and optimization protocol.

As shown in Table~\ref{tab:candidate_budget}, $N=8$ increases mean probe-measured target-accent accuracy by 2.49 pp and mean $\Delta p_{\mathrm{tgt}}$ by 1.93 pp relative to $N=1$.
The target-accent gains are positive for all three paired training seeds.
Mean WER increases by 0.29 pp, while SIM changes only from 0.7139 to 0.7134.
These results support the use of multiple stochastic realizations for constructing supervision with stronger downstream target-accent evidence under the selection probe $q$, while showing a small mean increase in WER.

\subsection{Alternative Supervision Strategies}
\label{sec:ablation}

\begin{table}[t]
\caption{
Alternative strategies for using feasible candidates.
Results are averaged over three training seeds.
}
\label{tab:ablation}
\centering
\small
\setlength{\tabcolsep}{1.60pt}
\begin{tabular}{lcccc}
\toprule
Strategy
& Acc. (\%) $\uparrow$
& $\Delta p_{\mathrm{tgt}}$ (pp) $\uparrow$
& WER (\%) $\downarrow$
& SIM $\uparrow$ \\
\midrule
Random Near-Best
& 29.82 & +10.90 & 8.18 & 0.7130 \\
Soft Top-2
& 29.88 & +10.56 & 7.82 & \textbf{0.7162} \\
Soft Near-Best
& 30.04 & +10.94 & 8.38 & 0.7156 \\
\textbf{Selected Top-1}
& \textbf{30.31} & \textbf{+11.01} & \textbf{7.81} & 0.7135 \\
\bottomrule
\end{tabular}
\end{table}

Table~\ref{tab:ablation} compares alternative ways of using feasible candidates.
Selected Top-1 uses the highest-ranked candidate.
Random Near-Best samples from candidates within $\epsilon=0.05$ of the highest target-accent probability.
Soft Top-2 combines losses from the two highest-ranked candidates, while Soft Near-Best applies the same soft-min objective to candidates within the same $\epsilon$ margin.

Selected Top-1 gives the highest mean target-accent accuracy and $\Delta p_{\mathrm{tgt}}$ and the lowest mean WER, while Soft Top-2 gives the highest mean SIM.
We therefore use Selected Top-1 in the final system.
\section{Conclusions}
\label{sec:conclusion}

We studied synthetic AC supervision as selection over multiple stochastic waveform--token realizations of the same source--accent condition.
Our framework first learns a categorical accent-conditioned generator, uses it to construct and assess multiple candidate conversions, and transfers the token sequence paired with a selected waveform to final autoregressive adaptation.
The resulting converter uses the source transcript and categorical target-accent label for token generation and requires neither target-accent reference speech nor multi-candidate selection at inference.
In the controlled comparison on the same source--accent conditions, eight-sample supervision improves probe-measured target-accent accuracy by 2.49 pp, with a 0.29-pp increase in mean WER and nearly unchanged speaker similarity.
Under the common output-level evaluation, the final system achieves the lowest WER among the evaluated systems and the highest observed mean ratings on all three perceptual measures.
The evaluation remains limited to six target accents and to external systems with different training and inference conditions.
Future work will extend the evaluation to broader accent settings and independently evaluate the matched multi-sample effect.

\section*{Acknowledgment}
This work was supported by the French Agence Nationale de la Recherche (ANR) through the SpeechPrivacy project (ANR-23-CE23-0022).
ChatGPT was used solely for language editing and stylistic refinement.
All scientific claims, experimental results, analyses, and conclusions were independently reviewed and verified by the authors.
\bibliographystyle{IEEEbib}
\bibliography{strings,refs}

\end{document}